# Research Funding in the Middle East and North Africa: Analyses of Acknowledgments in Scientific Publications indexed in the Web of Science (2008-2021)

Jamal El-Ouahi [1,2]

[1] Centre for Science and Technology Studies (CWTS), Leiden University, Leiden, Netherlands
j.el.ouahi@cwts.leidenuniv.nl
https://orcid.org/0000-0002-3458-7503

[2] Clarivate Analytics, Dubai Internet City, Dubai, United Arab Emirates

**Abstract**
Funding acknowledgments are important objects of study in the context of science funding. This study uses a mixed-methods approach to analyze the funding acknowledgments found in 2.3 million scientific publications published between 2008 and 2021 by authors affiliated with research institutions in the Middle East and North Africa (MENA). The aim is to identify the major funders, assess their contribution to national scientific publications, and gain insights into the funding mechanism in relation to collaboration and publication. Publication data from the Web of Science is examined to provide key insights about funding activities. Saudi Arabia and Qatar lead the region, as about half of their publications include acknowledgments to funding sources. Most MENA countries exhibit strong linkages with foreign agencies, mainly due to a high level of international collaboration. The distinction between domestic and international publications reveals some differences in terms of funding structures. For instance, Turkey and Iran are dominated by one or two major funders whereas a few other countries like Saudi Arabia showcase multiple funders. Iran and Kuwait are examples of countries where research is mainly funded by domestic funders. The government and academic sectors mainly fund scientific research in MENA whereas the industry sector plays little or no role in terms of research funding. Lastly, the qualitative analyses provide more context into the complex funding mechanism. The findings of this study contribute to a better understanding of the funding structure in MENA countries and provide insights to funders and research managers to evaluate the funding landscape.

**Keywords**
Science Funding, Funding acknowledgment, Research System, Middle East and North Africa.

## 1. Introduction

Funding organizations play a significant role in the advancement of science (Braun, 1998). In addition to funding provided to public universities and research institutions, they also fund researchers on specific and selective programs. Latour and Woolgar (1986) argue that scientific facts are not simply "discovered" but actively constructed through social processes in the laboratory. Funding acknowledgments in scientific papers are part of these processes. They give credit to individuals and organizations that contributed to the research. They are also used for self-promotion by researchers to show their connections to prestigious organizations, influential colleagues and well-funded projects (Latour & Woolgar, 1986). Analyzing funding acknowledgments can reveal hidden networks of scientific collaborations, social dynamics within and between research teams but also the social context and the dynamics that shaped the research reported in scientific publications (Cronin & Weaver, 1995). Since 2008, funding acknowledgments found in scientific publications have been captured in the Web of Science (Clarivate, 2022). Such data enables researchers to conduct analyses on funding sources (Alvarez-Bornstein & Montesi, 2021; Giles & Councill, 2004; Paul-Hus et al., 2016). Early research on funding acknowledgments analyzed the coverage by country, disciplines, and document types as well as methods to unify funding organizations (Álvarez-Bornstein et al., 2017; Costas & van Leeuwen, 2012; Sirtes & Riechert, 2014). Later, research analyzing funding acknowledgments focused on funding mechanisms. Möller (2019) identified the major funders and their contribution to the nationwide performance in five European countries. Wang et al. (2012) compared the research funding systems in ten countries and found that China is dominated by a single funder whereas the United Kingdom has more funding sources. Other studies focused on specific groups of countries such as the G9 countries (Huang & Huang, 2018), the Global South (Chankseliani, 2023), the BRICS (Shueb & Gul, 2023), or specific countries (Álvarez-Bornstein et al., 2018; Alvarez & Caregnato, 2018, 2021; Costas & Yegros-Yegros, 2013; Díaz-Faes & Bordons, 2014; Gao et al., 2019; Gök et al., 2016). More recently, Chataway et al. (2019) analyzed the trends in science funding support in Sub-Saharan Africa and found that the levels of funding and cross-country engagement were low and that there was a need for capacity building. They also noted that there was a growing interest towards science funding in Sub-Saharan Africa at the national, regional, and international levels.

Traditionally, economies in MENA have been dependent and centralized around natural resources (World Bank, 2019). In the Middle East and North Africa (MENA), nations have been looking to become knowledge-based economies (OECD, 1996). For example, oil-exporting countries established research funds and new universities as part of their transition to knowledge economies (Currie-Alder, 2019). A report from UNESCO (2015) provides some insights into the funding of science in MENA. At that time, the investment in Research and Development reported by UNESCO was reported as low in the region. However, some changes were already taking place in terms of funding allocation, priority setting and promotion of collaboration to address large-scale societal challenges (Currie-Alder, 2019). Considerable investments in science and technology capacity have been made recently in MENA countries to promote research and innovation (Schmoch et al., 2016; Shin et al., 2012; Siddiqi et al., 2016). These investments resulted in the expansion of research funds over the past two decades in several Arabic-speaking countries (Currie-Alder, 2015; Currie-Alder et al., 2018) with research assessments privileging "collaboration with distant, scientifically proficient partners abroad, in order to connect with global networks and rise in international rankings of academic quality".

To the best of my knowledge, no analysis of funding acknowledgments found in scientific publications has been conducted to better understand the structure and the trends of scientific funding for this specific region. This study aligns with a larger framework of scholarly communication where acknowledgments are used along authorship and citations as the "reward triangle" (Cronin & Weaver, 1995). This framework allows "a more nuanced understanding of scholarly communication and interaction" (Cronin, 1995). Analyses of funding acknowledgments reveal the broader research science ecosystem that sustains research activities. In this broader ecosystem, Cronin's concept of "subauthors" encompasses the significant contributions and roles played by people and organizations in both the development of research and the writing of scientific papers (Cronin et al., 2003). Acknowledgments have also been frequently considered as "super citations" (Cronin et al., 1993). Acknowledgments and citations show a high level of cultural consensus, and both describe networks of interactions and influence (Cronin & Weaver, 1995). Since funding acknowledgments "provide a revealing window onto trends in collaboration beyond co-authorship" (Cronin & Weaver, 1995), their study may also inform research managers and policy makers by guiding funding strategies or by unveiling hidden synergies and influence that cannot be identified through regular co-authorship or citation analyses.

In the context of recent investments made in science and technology in MENA countries, this study compares the contribution of their research funders from the research publications perspective. This article explores the structure and the recent trends in science funding in MENA with reflections at the regional and national levels. This study also aims to better understand the funding mechanism in science with regard to collaboration and publishing.

More specifically, in this study, I address the following research questions:
- To what extent has the funding structure evolved over the past few years in MENA?
- What are the characteristics of the major funders in MENA, in terms of type and location?

These aspects provide insights into the landscape of funders in MENA. These insights are also particularly helpful for policymakers to better understand the funding structures in various national science systems.

This paper is organized as follows. Section 2 describes the data and the methods used to analyze the funding acknowledgments in the MENA region. Then, the results of the analyses are presented in Section 3. Finally, the findings of this study are briefly discussed in Section 4.

## 2. Data and Methods

### 2.1. Geographic coverage

According to the World Bank (2019), MENA consists of the following countries: Algeria, Bahrain, Djibouti, Egypt, Iran, Iraq, Jordan, Kuwait, Lebanon, Libya, Morocco, Oman, Palestine, Qatar, Saudi Arabia (KSA), Syria, Tunisia, the United Arab Emirates (UAE) and Yemen. Pakistan, Afghanistan, and Turkey are also considered in this study as they are commonly included in the MENA region (MENAP and MENAT).

### 2.2. Data source

This analysis is based on the scientific publications indexed in the Web of Science Core Collection (WoS) which contains greater coverage of funding acknowledgments than other data sources such as Scopus or PubMed (Kokol & Vošner, 2018; Liu, 2020; Mugabushaka et al., 2022; Sterligov et al., 2020). All document types published between 2008 and 2021 with at least one author affiliated with an institution in MENA and indexed in one of the citation indexes of WoS are considered in this study. 2.3M publications satisfy these criteria as of 31 October 2022.

Table 1 lists the number of publications (P) indexed in WoS for all the countries in MENA between 2008 and 2021 along with the proportion of international co-authorship (%Int). Countries with similar output levels in terms of scientific publications are shaded in Table 1 with the same color.

As also shown in Table 1, it is worth reminding that these countries have different sizes in terms of population, Gross Domestic Product (GDP), Research and Development Expenditure (RDE as % of GDP) (World Bank, 2023), and funding structures that we aim to better understand in this study.

**Table 1. Population size, GDP, RDE and number of publications in WoS for each country in MENA (2008-2021)**

| Country | ISO | 2022 Population (Thousands) | GDP (current US$ in millions) [1] | RDE [1] | P | %Int |
|---|---|---|---|---|---|---|
| *WORLD* | | 7,950,000 | | | *40.7M* | 19% |
| *MENA* | | 845,125 | | | *2.3M* | 35% |
| Turkey | TUR | 84,980 | 907,118 (2022) | 1.40% (2021) | 661,839 | 24% |
| Iran | IRN | 88,551 | 413,493 (2022) | 0.79% (2019) | 615,666 | 20% |
| Saudi Arabia | SAU | 36,409 | 1,108,572 (2022) | 0.45% (2021) | 262,565 | 70% |
| Egypt | EGY | 110,990 | 476,748 (2022) | 0.91% (2021) | 238,118 | 50% |
| Pakistan | PAK | 235,825 | 374,697 (2022) | 0.16% (2021) | 216,056 | 50% |
| Tunisia | TUN | 12,356 | 46,304 (2022) | 0.75% (2019) | 91,189 | 68% |
| Algeria | DZA | 44,903 | 194,998 (2022) | 0.53% (2017) | 70,797 | 49% |
| UAE | ARE | 9,441 | 507,064 (2022) | 1.50% (2021) | 68,324 | 76% |
| Morocco | MAR | 37,458 | 130,913 (2022) | 0.66% (2010) | 63,443 | 43% |
| Iraq | IRQ | 44,496 | 264,182 (2022) | 0.04% (2021) | 42,423 | 49% |
| Jordan | JOR | 11,286 | 48,653 (2022) | 0.70% (2016) | 41,588 | 53% |
| Qatar | QAT | 2,695 | 236,258 (2022) | 0.68% (2021) | 40,102 | 50% |
| Lebanon | LBN | 5,490 | 23,132 (2021) | NA | 36,110 | 61% |
| Kuwait | KWT | 4,269 | 175,363 (2022) | 0.19% (2020) | 20,892 | 57% |
| Oman | OMN | 4,576 | 114,667 (2022) | 0.29% (2021) | 19,162 | 68% |
| Bahrain | BHR | 1,472 | 44,383 (2022) | 0.10% (2014) | 7,175 | 85% |
| Palestine | PSE | 5,044 | 19,112 (2022) | NA | 6,471 | 66% |
| Syria | SYR | 22,125 | 8,970 (2021) | 0.02% (2015) | 6,410 | 75% |
| Libya | LBY | 6,812 | 45,752 (2022) | NA | 5,963 | 54% |
| Yemen | YEM | 33,697 | 21,606 (2018) | NA | 3,510 | 59% |
| Afghanistan | AFG | 41,129 | 14,266 (2021) | NA | 2,089 | 79% |
| Djibouti | DJI | 1,121 | 3,515 (2022) | NA | 256 | 92% |

## 2.3. Data unification

WoS contains four fields with funding information (Clarivate, 2023):
1) Funding Text (FT): funding acknowledgments of the authors.
2) Funding Details (FD): descriptive grant data acquired from external grant repositories.
3) Grant Number (FG)
4) Funder (FO): name of the funding organization.

This study is based on the Funder names found in the Funding Text. Out of the 2.3M publications in the dataset under study, close to 1.6M (68%) do not contain funding data. A semi-automated process with human and manual quality checks was used to unify the non-unified FO names found in the remaining publications. This process is represented in Figure 1.

---

[1] Data is extracted from the latest year indicated within the brackets in Table 1, or denoted as NA if the information is not available.

There are many different variant names for each funding organization with many linked to only one or two publications. During this unification process, the objective was to identify and unify all the funders for each MENA country which contribute to at least 1% of the publications at the national level.

The first step of this unification process consists of extracting the FO variant names from the publications of the dataset under study. The frequency of occurrence is calculated for each variant. Next, the FO variant names are preprocessed by removing duplicates resulting in 41,182 unique variant names. A key step in the unification process is to create a controlled vocabulary of variant names by considering the most frequently used ones as proposed by Wang and Shapira (2011) and Sirtes (2013). The most frequent variants are then clustered by using fuzzy matching techniques provided by the Dataiku platform (Dataiku, 2023). 5,242 initial clusters are obtained as a result. These clusters are manually verified, and the incorrectly matched variant names are excluded from their initial clusters. The fuzzy matching algorithm is then reapplied on the labels of the obtained clusters, until no more clusters are created. The clustering is manually checked after each iteration.

The full names of the funders, found by searching the labels of the clusters on the web, were preferred as the unified names over their acronyms to avoid confusion between funders sharing the same acronyms across different countries. For example: "Ministry of Higher Education and Scientific Research" was preferred over "MOHESR". When possible, based on the funding text, the country of origin of the funder is added for clarity to such ambiguous names. Also, in many cases, authors only mention the names of the funding programs in their acknowledgments. In such instances, the program names were searched on the Internet and linked to the preferred name of the corresponding FO in the controlled vocabulary.

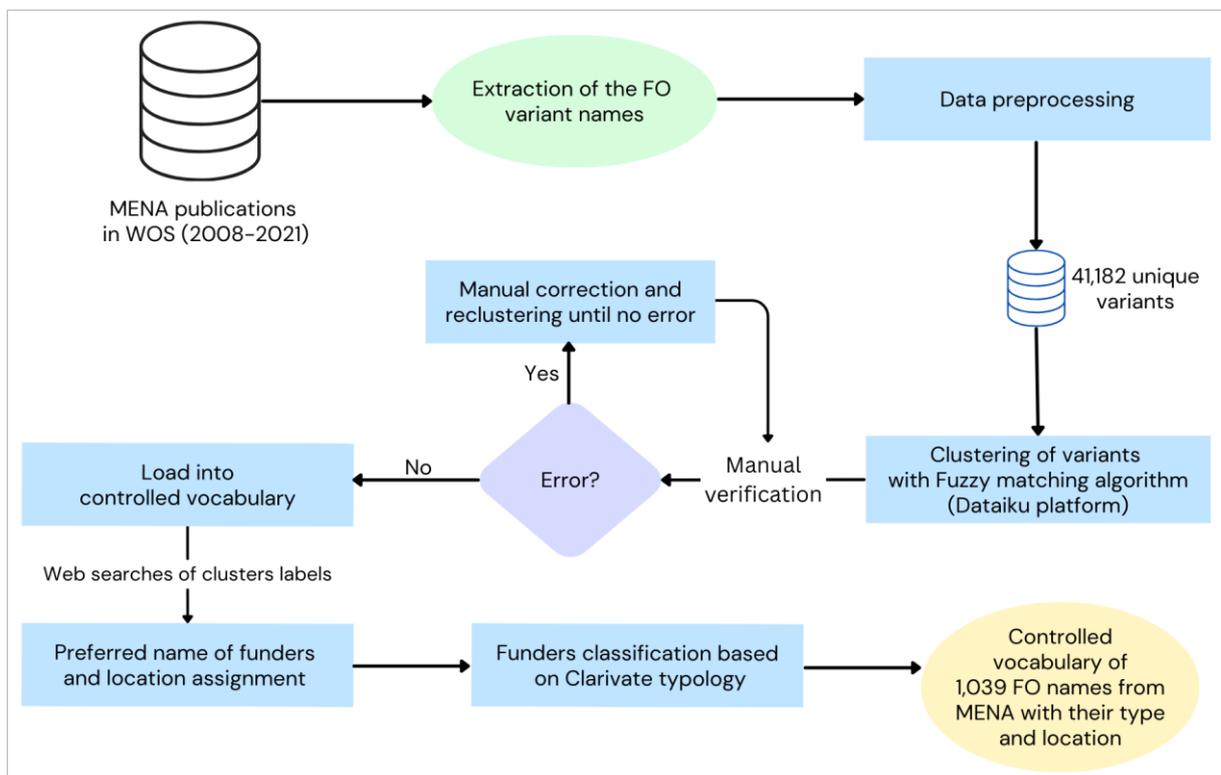

**Figure 1: Semi-automated process of the standardization of funders names**

For each unified name a country was assigned. The country of the funder is not always mentioned in the acknowledgments. Based on the funder's name, a search on the internet allowed to find the relevant country of the funder. This information is used in the analyses to classify a funder as either *domestic* or *foreign*, from the perspective of the country of the related publications, but also as a funder located in or outside the MENA region. It is worth noting that some funders or programs are international. In such case, the relevant region or group of countries' names were assigned (e.g. Europe, Arabia, South Asia…).

Then, a type was assigned to each funder by using the existing classification of funders defined by Clarivate and available in WoS and InCites (Clarivate, 2023): Academic, Academic System, Corporate, Global Corporate, Government, Health, National Academy, Nonprofit, Partnership, Research Council, and Research Institute. In most cases, applying this typology to funders was done as follows. For instance, if the name of a funder contained the string "Universi", then the *Academic* type was assigned to this agency. Similarly, when "Minist" was found, the *Government* type was assigned. If "Hospital" or "hopit" was found in the agency name, the funder was classified as *Health*. Similarly, "council", "conseil", "conselho", "consejo" were searched to assign the relevant type. Names of funders containing "Assoc" or "Foundation" or "Fondation" were classified as *Nonprofit* unless related to the government. Other strings such as "Inc", "Corp", "GMBH", "LLC", "Co Ltd" were evidence of a *Corporate* type, with the *Global Corporate* type defined as a company that operates in two or more countries. For the rest, similarly to the location assignment, web searches were used to assign the relevant type to the remaining funders.

As a simple example, "University Mohammed VI Polytechnic" is the official name of a university located in Morocco acknowledged as a funder and the preferred name of the following variant names:
- Mohammed VI Polytechnic University
- Mohammed VI Polytechnic University at Ben Guerir Morocco
- Mohammed VI Polytechnic University de Benguerir Morocco
- Mohammed VI Polytechnic University Morocco
- Mohammed VI Polytechnic University of Benguerir
- Mohammed VI Polytechnic University of Benguerir Morocco
- Mohammed VI Polytechnic University UM6P of Ben Guerir Morocco
- Mohammed VI Polytechnic University UM6P of Benguerir Morocco
- University of Mohammed VI Polytechnic
- University of Mohammed VI Polytechnic Ben Guerir Morocco
- Welcome Grant From The Mohammed VI Polytechnic University UM6P of Ben Guerir Morocco
- Welcome Grant of The Mohammed VI Polytechnic University UM6P
- Mohammed VI Polytechnic University UM6P
- University Mohammed VI Polytechnic

The following variant names could not be unified nor assigned to a single funder and a single country as there are multiple entities with the same name located in different countries around the world:
- Deanship of the Scientific Research
- Department of Obstetrics and Gynecology
- Department of Scientific Research

- Ministry of Public Health
- Ministry of Science and Education
- University of Technology

These variants were not included in the dataset under study.

### 2.4. Unified funders in MENA

Table 2 lists the number of unified funders in MENA before and after the unification process along with the share they represent across the region. 1,254 names of funders from 69 countries were already unified in WoS as of 31 October 2022, with 236 of them located in 8 MENA countries, mainly in Turkey and to a lesser extent in Egypt. After the unification, the dataset under study contains 1,039 unified names of funders from the 22 MENA countries as of 16 March 2023. The list of unified funders with their type and country is available in a data repository (El-Ouahi, 2023).

**Table 2. Number and percentage of unified funders in MENA before and after the unification process.**

|  | Before Unification *31 October 2022* | | After Unification *16 March 2023* | |
|---|---|---|---|---|
| **Country** | **Number** | **Percentage** | **Number** | **Percentage** |
| *MENA* | *236* | *100%* | *1,039* | *100%* |
| Turkey | 194 | 82% | 240 | 23% |
| Egypt | 28 | 12% | 61 | 6% |
| Iran | 4 | 2% | 307 | 30% |
| Saudi Arabia | 4 | 2% | 59 | 6% |
| UAE | 2 | 1% | 29 | 3% |
| Oman | 2 | 0.8% | 15 | 2% |
| Pakistan | 1 | 0.4% | 100 | 10% |
| Kuwait | 1 | 0.4% | 8 | 1% |
| Morocco |  |  | 42 | 4% |
| Algeria |  |  | 36 | 3% |
| Tunisia |  |  | 29 | 3% |
| Jordan |  |  | 25 | 2% |
| Iraq |  |  | 22 | 2% |
| Lebanon |  |  | 16 | 2% |
| Qatar |  |  | 10 | 1% |
| Syria |  |  | 9 | 1% |
| Yemen |  |  | 7 | 1% |
| *Regional* |  |  | 6 | 0% |
| Libya |  |  | 5 | 0% |
| Palestine |  |  | 4 | 0% |
| Bahrain |  |  | 4 | 0% |
| Djibouti |  |  | 3 | 0% |
| Afghanistan |  |  | 3 | 0% |

After the unification process, the unified names of the funders cover now all the MENA countries with Iran and Turkey dominating the region in terms of number of funders, followed by Pakistan, Saudi Arabia and Egypt. The number of funders varies between countries, reflecting differences in the size of countries in terms of funding capacity or publication level (as shown in Table 1) and also in funding structures.

**2.5. Counting method and proportion of papers with funding acknowledgments**

Various analyses were conducted using WoS data. A full counting method was used in this study, i.e., co-authored publications at the country level are fully counted as a publication of each country. Similarly, publications with acknowledgments of multiple funders at the country level were fully counted for each funder.

The number of funders and their contribution to each country's publications depends also on the presence of foreign funding organizations which have provided funds to international co-authors. The foreign funding agencies acknowledged in scientific publications of MENA countries might have been acknowledged by international co-authors who collaborated with researchers based in MENA. Hence, in this study, an important distinction is made between the contribution of funders to international publications (IP, i.e., a country publication with at least one international co-author) and the contribution the funders have when considering only domestic publications (DP, i.e., a country publication without international co-authors).

The proportion of papers with funding acknowledgments (PWFAs) for a given country is defined as the number of papers from this country with funding acknowledgments divided by the total number of papers counted for that same country. The proportion of papers with funding acknowledgments is calculated for domestic publications (%DPWFA) and for international publications (%IPWFA).

**2.6. Limitations**

The analysis of funding acknowledgments is a complex task because such data is usually included in a separate section of a scientific paper in a non-structured way along with other types of acknowledgments (Cronin, 1995). Thus, analyzing funding acknowledgments from raw data and unifying the names of funders with a proper typology is not a straightforward task. Such analysis comes with certain limitations that one needs to be aware of when analyzing funding acknowledgments (Aagaard et al., 2021; Paul-Hus et al., 2016; Sirtes & Riechert, 2014; Tang et al., 2017; Wang & Shapira, 2015). There might be a certain impact of cultural factors on what authors acknowledge in their papers (Rigby, 2011); funding acknowledgments may focus on external funding bodies and may tend to significantly ignore internal resources received from the researchers' own institutions. It has been shown that authors do not always acknowledge their funders (Costas & Yegros-Yegros, 2013) and that acknowledgment practices vary depending on research areas (Grassano et al., 2017). Hence, the analysis of funding acknowledgments may offer only a partial view on the landscape of funding organizations. Last, this study is limited to scientific papers published only in journals indexed in Web of Science. Álvarez-Bornstein et al. (2017) explored the completeness and accuracy of funding acknowledgments in the Web of Science and found that some funding information was lost in 12% of the articles of their dataset. Figure 2 shows the proportion of publications with funding acknowledgments that present complete funding information (FO, FT and FG) in the current study's dataset.

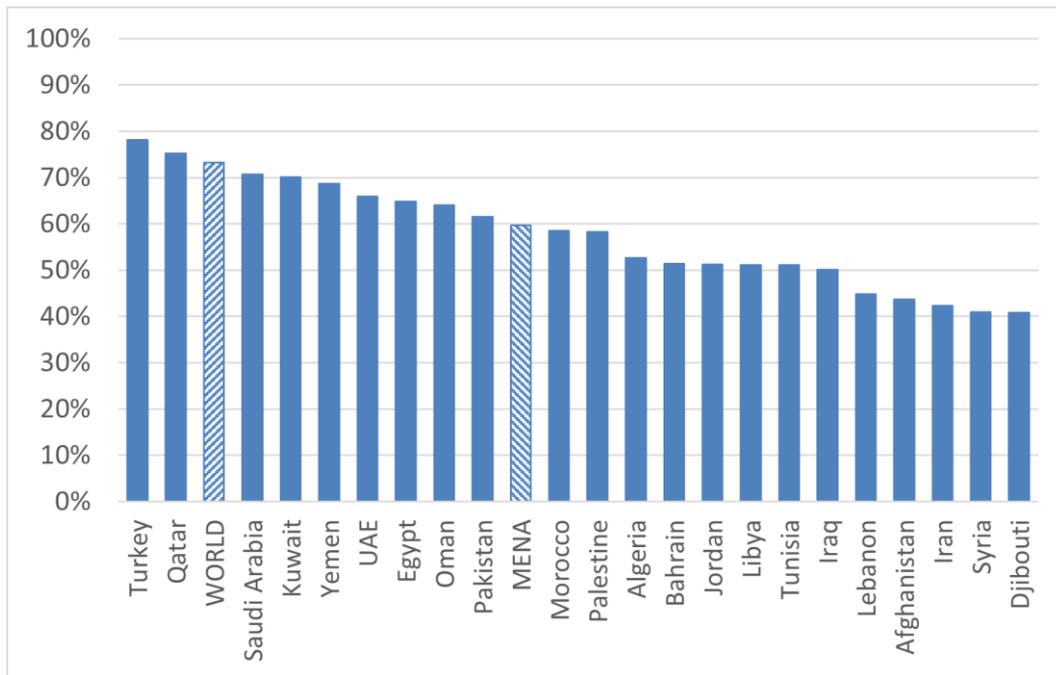

**Figure 2: Percentage of PWFAs with complete funding information (FO, FT and FG) in MENA countries and the World (2008-2021)**

About 60% of the PWFAs from MENA published between 2008 and 2021 present complete funding information, i.e., the publications contain the FO, FT and FG information in their related records. This proportion varies by country. Turkey, Qatar and Saudi Arabia lead the region with more than 70% of their PWFAs containing complete funding information. By contrast, PWFAs from Djibouti, Syria and Iran show a proportion of about 40% of records with complete funding information.

All these caveats must be kept in mind when analyzing and interpreting funding data. Also, the specific funding level or the project types supported by the funders are not considered in this study. Despite all these limitations, the funding acknowledgments available in WoS provide an important information resource to answer questions related to research funding. More specifically, this data offers some opportunities to explore the source of the funding received by researchers in terms of location and type of funder. Also, the large scale of data availability allows for country comparisons.

## 3. Findings

### 3.1. Proportion of PWFAs by country

Before analyzing the unified funders, the proportion of papers with funding acknowledgments (PWFAs) is examined for each country in MENA. These proportions are shown in Figure 3, which also includes the share of PWFAs at the MENA and World levels as benchmarks. Here a distinction is made between the international publications of a country and its domestic publications, i.e., publications without international co-authors.

When looking at the international publications, the proportion of PWFAs is 61% for the World and 49% for the MENA region. Figure 3 shows that Saudi Arabia, Qatar, Pakistan and Djibouti lead the region in terms of proportion of IPWFAs. It is worth reminding that the countries in MENA have different levels of scientific output. For example, Yemen, Palestine, Libya, Syria, Bahrain, Afghanistan and Djibouti are countries with fewer than 7,500 WoS-indexed papers published between 2008 and 2021 and with a relatively high level of international co-authorship, as shown in Table 1. Djibouti counts the lowest output with only 256 publications during the same period and has an international co-authorship rate of 92%. These countries show a much lower proportion of PWFAs for their domestic publications than for all their international publications. This suggests that a relatively high share of funders found in PWFAs of such countries might have been acknowledged by international co-authors who collaborated with researchers from these specific countries. In contrast, some other countries such as Iran, Kuwait show small differences in terms of proportions of PWFAs when comparing their publications with and without international co-authors.

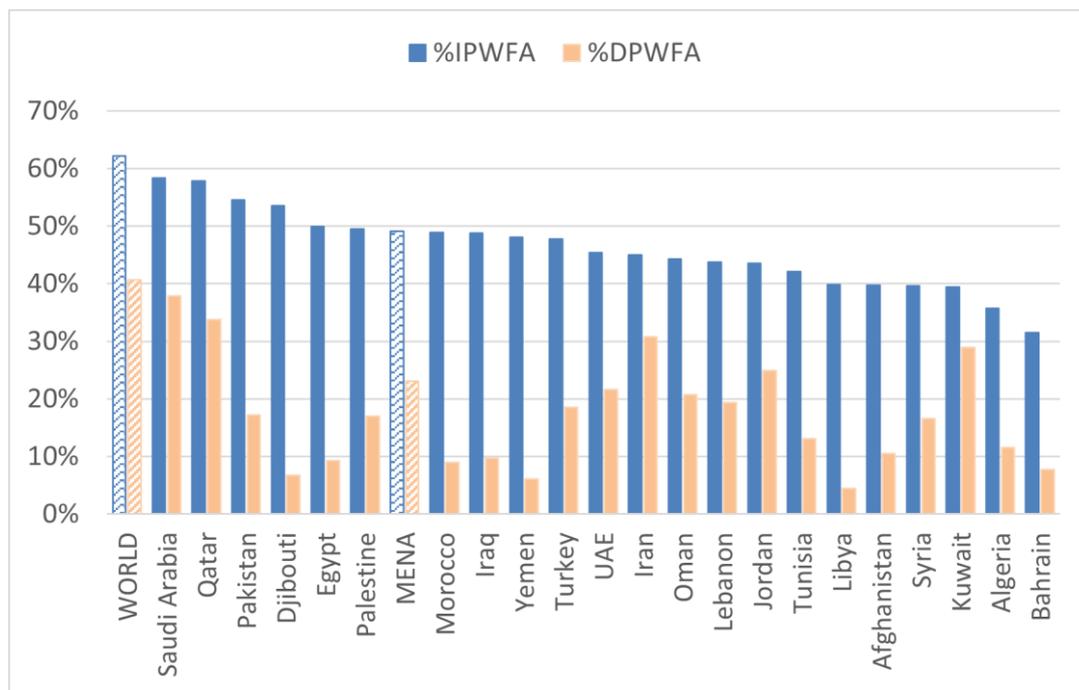

**Figure 3: Percentage of PWFAs for international (%IPWFA) and domestic publications (%DPWFA) in MENA countries and the World (2008-2021)**

The trend of the proportion of the PWFAs for each MENA country over the study period is shown in Figure 4, with a distinction between the international publications of a country (%IPWFA) and its domestic publications (%DPWFA). The trends for the World and the MENA region are also included as benchmarks with dotted lines. This figure provides some additional insights into the evolution of the funding of scientific publications in MENA.

One can notice the upward trends across most countries as well as in the World and MENA. These uptrends of the proportion of PWFAs can be observed in the international publications as well as in the domestic publications of each country. It is worth noting that the MENA countries as well as the world started from a relatively low level of PWFAs in 2008 (approximately 10% on average), which might be due to low coverage of funding acknowledgments in publications in WoS in 2008. Another reason could be a lack of funding acknowledgments made by researchers in their publications in the early years of the study period. Acknowledging funders explicitly in scientific publications might have been less common in the past than it is currently.

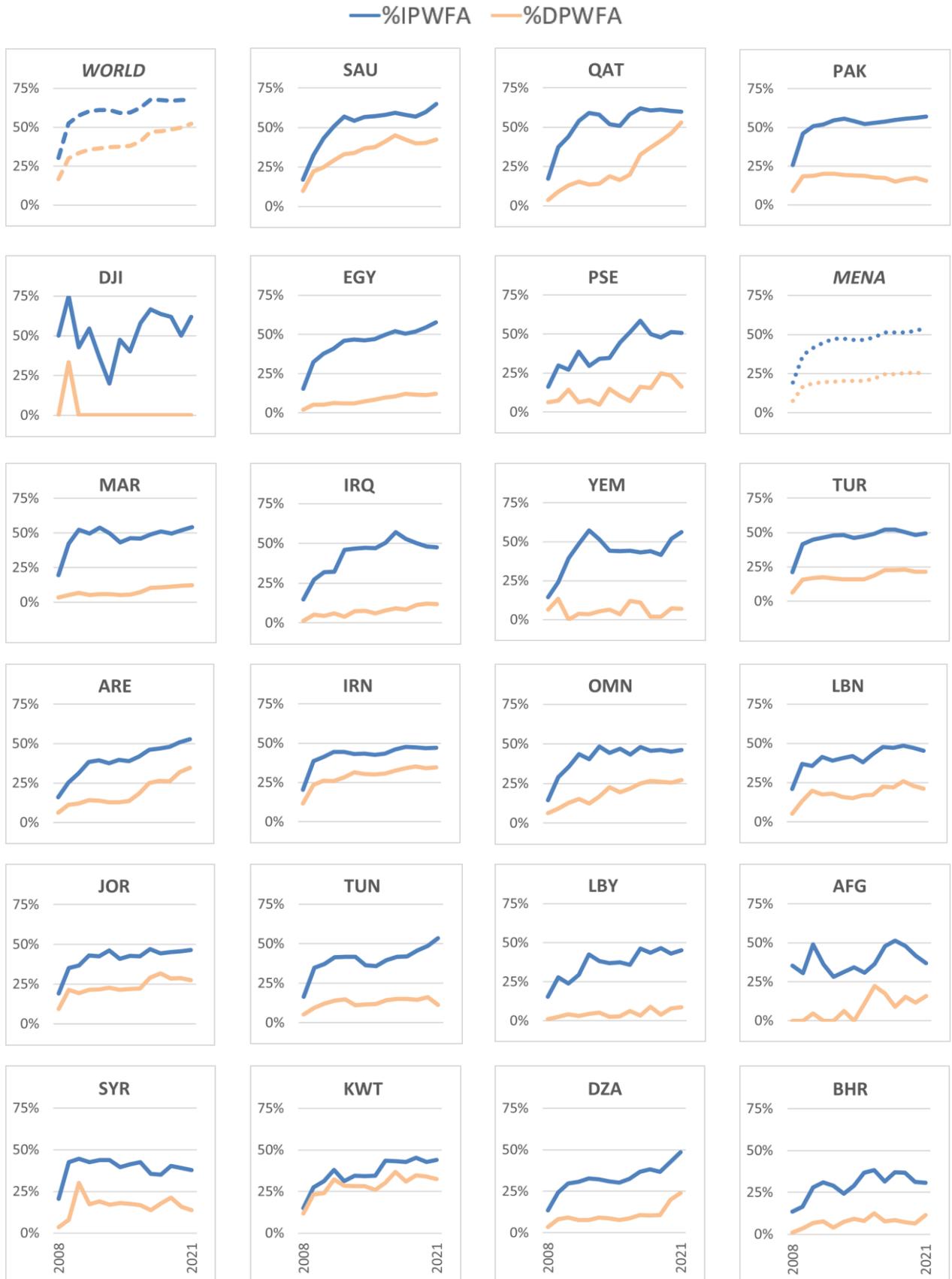

**Figure 4. Trends of the percentage of publications with funding acknowledgments for domestic and international publications in MENA countries and the World (2008-2021)**

The countries are shown in Figure 4 from the highest average proportion of PWFAs between 2008 and 2021 (top left) to the lowest (bottom right) when considering the international publications of the country. Saudi Arabia, Qatar and Pakistan lead the region. The high growth of the proportion of PWFAs in Saudi Arabia, Qatar and the UAE may also reflect the investment in research and funding policies recently set in these countries. In the early 2000s, according to its *Seventh Development Plan,* Saudi Arabia increased its total amount of spending on research and development activities with the goal to reach 2% of its GDP by 2025 (Kingdom of Saudi Arabia Ministry of Economy and Planning, 2000). This level of investment in research brought Saudi Arabia from being the 50$^{th}$ largest spender in the world in 2009 to the 16$^{th}$ rank in 2016; one example of such spending on research in Saudi Arabia consisted of the establishment of King Abdullah University of Science and Technology (KAUST) in 2009, which currently has a 20-billion-dollar endowment (Research Professional, 2020). This increase is also highlighted in the recent Saudi National Transformation Program 2020 (Saudi Arabia's Vision 2030, 2020). Similarly, the Ministry of Education in the UAE set a target of 1.5% of its GDP for 2021 for the expenditure on research and development. In addition to internal funding for research provided by Higher Education Institutions in the UAE, a few national sources of funding are available, such as the Abu Dhabi Department of Education and Knowledge Award, the Sheikh Hamdan Award for Medical Sciences and more recently the National Fund or *Sandooq Al Watan* (Al Marzouqi et al., 2019). In Qatar, the Qatar National Research Fund (QNRF) was established in 2006 as a national organization with the mission to fund and promote research in Qatar as well as scientific research cooperation. As a result of such development, QNRF awarded $650 million in grants in the first 7 years of its existence (Weber, 2014).

For most countries, the trends in the proportion of PWFAs for domestic and international publications are similar, but some countries stand out. For instance, the differences between the proportion of PWFAs in international and domestic publications is relatively small for Iran, and Kuwait throughout the study period and for Qatar in recent years. The differences between the share of PWFAs in domestic and international publications are relatively constant for several countries such as Saudi Arabia, Lebanon, Jordan, Oman, United Arab Emirates, Morocco, Algeria and Bahrain. However, these differences are increasing for Yemen, Libya, Egypt and more specifically Pakistan and Tunisia, which both show a slight decrease of the share of PWFAs in its domestic publications in recent years.

Syria, and Afghanistan show downtrends in the share of PWFAs in recent years, which might be due to the recent conflicts or unrest in these countries. Figure 4 also reveals a decrease of the share of PWFAs around 2013 for some countries, such as Tunisia, Libya, or Yemen. It is unclear what caused these decreases. One reason could be the so-called *Arab Spring* which impacted several countries in the region. Then, the share of PWFAs increased in these countries in the following years. Science systems are particularly vulnerable to wars, social and political unrests, which limit the availability of funding granted to scientists or may even damage the national science systems.

### 3.2. Major funders in MENA

The unified data allows the identification of the major funders for each MENA country. Table 3 lists the number of major agencies for each country in MENA with the distinction between the publications of the country with and without international co-authors. Major funders are defined as agencies which contribute to at least 1% of a country's total number of publications with and without international co-authors indexed in WoS.

**Table 3. Number of major funders contributing to at least 1% of each country's total number of publications in WoS for each country in MENA (2008-2021)**

| Country | Number of Major Funders | | Country | Number of Major Funders | |
|---|---|---|---|---|---|
| | *International Publications* | *Domestic Publications* | | *International Publications* | *Domestic Publications* |
| Palestine | 109 | 3 | Jordan | 10 | 5 |
| Qatar | 97 | 4 | Oman | 11 | 3 |
| Morocco | 98 | 1 | Pakistan | 44 | 1 |
| Djibouti | 21 | 1 | Yemen | 10 | - |
| Afghanistan | 10 | 2 | Iran | 16 | 4 |
| Saudi Arabia | 18 | 6 | Iraq | 13 | 2 |
| Egypt | 23 | 2 | Libya | 8 | - |
| Lebanon | 13 | 5 | Tunisia | 6 | 1 |
| Kuwait | 12 | 4 | Algeria | 9 | 1 |
| UAE | 17 | 4 | Bahrain | 9 | 1 |
| Syria | 12 | 2 | Turkey | 47 | 2 |

These results suggest that MENA countries seem to have different structures of research funding systems. Several dozen funders appear in at least 1% of the national number of international publications in countries such as Palestine, Qatar and Morocco. It appears that these countries have more diversified funding sources than countries dominated by one or two major agencies like Algeria, Tunisia, Libya, and Bahrain. Other countries such as Djibouti, Afghanistan, Saudi Arabia, or Egypt show a few funders. However, when considering only domestic publications, there is a clear decrease in the number of funders for all the countries in MENA. This decrease is particularly greater for the countries with a relatively high rate of international co-authorship. Table 3 also allows to make a distinction between countries in MENA with only one or more funders based on their contributions to domestic publications. Saudi Arabia shows the highest number of funders acknowledged in domestic publications than any other country in MENA. Then Jordan, Lebanon, Iran, Qatar and the UAE follow. Then, several countries such as Morocco, Algeria, Tunisia or Pakistan count only one major funder contributing to domestic publications.

Overall, Table 3 provides a snapshot of the number of funders in the MENA countries. It provides some useful information about the potential funding support available for each country.

To better understand the structure of research funding in MENA, when available, Table 4 lists the three main funders for each country (CU) in MENA based on the related number of domestic publications with funding acknowledgments (DPWFAs), the proportion these publications represent over the total number of domestic publications (%DP) over the study period, the country of the funder (FCU) and its type.

Unsurprisingly, Table 4 lists almost exclusively domestic funders, except for Afghanistan, Djibouti and Syria. For some countries like Algeria, Morocco, Tunisia, Bahrain, and Turkey it was not possible to find three funders because they are dominated by a single funder. Then, the countries dominated by double funders include Iran, Iraq, Libya, and Tunisia. The academic sector is the most prominent contributor across MENA with 24 universities contributing more than 1% of the total number of domestic publications in their countries. Then, the government institutions, mainly ministries of higher education and research, national research centers, foundations, or funds. Libya and Yemen are not listed in Table 5 since there is no major funder acknowledged in their domestic publications.

**Table 4. Top 3 funders for each country by number of domestic publications with funding acknowledgments in WoS for MENA countries (2008-2021)**

| CU | FO | DPWFAs | %DP | FCU | Type |
|---|---|---|---|---|---|
| AFG | World Health Organization | 8 | 2.5 | World | Nonprofit |
| | Khatam Al Nabieen University | 6 | 1.9 | AFG | Academic |
| ARE | United Arab Emirates University | 1,256 | 6 | | Academic |
| | Khalifa University of Science and Technology | 873 | 4.2 | ARE | Academic |
| | University of Sharjah | 372 | 1.8 | | Academic |
| BHR | Arabian Gulf University | 81 | 3 | | Academic |
| | University of Bahrain | 63 | 2.3 | BHR | Academic |
| DJI | Armed Forces Health Surveillance Center | 1 | 6.7 | USA | Government |
| DZA | Ministry of Higher Education and Scientific Research, Algeria | 2,351 | 6.6 | DZA | Government |
| EGY | Science and Technology Development Fund | 1,808 | 1.6 | | Government |
| | National Research Centre, Egypt | 1,651 | 1.4 | EGY | Research Institute |
| IRN | Islamic Azad University | 12,708 | 2.7 | | Academic |
| | Iran National Science Foundation | 10,301 | 2.2 | IRN | Government |
| | Shiraz University of Medical Sciences | 5,154 | 1.1 | | Academic |
| IRQ | Al Mustansiriyah University | 365 | 1.8 | | Academic |
| | Ministry of Higher Education and Scientific Research of Iraq | 333 | 1.6 | IRQ | Government |
| JOR | Jordan University of Science and Technology | 1,302 | 6.8 | | Academic |
| | University of Jordan | 1,037 | 5.4 | JOR | Academic |
| | Hashemite University | 263 | 1.4 | | Academic |
| SAU | King Saud University | 7,582 | 10.4 | | Academic |
| | King Fahd University of Petroleum Minerals | 4,112 | 5.7 | SAU | Academic |
| | King Abdullah University of Science & Technology | 2,958 | 4.1 | | Academic |
| KWT | Kuwait University | 1,655 | 18.7 | | Academic |
| | Kuwait Foundation for the Advancement of Sciences | 600 | 6.8 | KWT | Nonprofit |
| | Kuwait Institute for Scientific Research | 323 | 3.7 | | Research Institute |
| LBN | American University of Beirut | 1,209 | 8.8 | | Academic |
| | Lebanese National Council for Scientific Research | 501 | 3.7 | LBN | Research Council |
| | Lebanese University | 177 | 1.3 | | Academic |
| MAR | Centre National pour la Recherche Scientifique et Technique | 1,003 | 2.8 | MAR | Government |
| OMN | Sultan Qaboos University | 655 | 11.2 | | Academic |
| | The Research Council Oman | 275 | 4.7 | OMN | Research Council |
| | Ministry of Higher Education, Research & Innovation, Oman | 155 | 2.7 | | Government |
| PAK | Higher Education Commission of Pakistan | 10,298 | 9.7 | PAK | Government |
| PSE | An Najah University Palestine | 98 | 4.6 | | Academic |
| | Birzeit University | 42 | 2 | PSE | Academic |
| | Palestine Technical University | 24 | 1.1 | | Academic |
| QAT | Qatar National Research Fund | 1,657 | 18.1 | | Nonprofit |
| | Qatar Foundation | 852 | 9.3 | QAT | Nonprofit |
| | Hamad Bin Khalifa University | 117 | 1.3 | | Academic |
| SYR | Atomic Energy Commission | 234 | 8.1 | World | Government |
| | Damascus University | 123 | 4.3 | SYR | Academic |
| TUN | Ministry of Higher Education and Scientific Research, Tunisia | 4,165 | 9 | TUN | Government |
| TUR | Turkiye Bilimsel ve Teknolojik Arastirma Kurumu | 33,293 | 6.3 | TUR | Government |

When available, the top three funders of international publications are listed in Table 5 along with the number of publications with funding acknowledgments (IPWFAs), the related proportion (%IP) for the corresponding country of publication (CU) over the study period, the country of the funder (FCU) and its type. This table also provides some information in terms of source of funding support received. For most countries in MENA, the funders are government organizations or organizations related to government entities. One can also see that, for some countries, the academic sector appears as the major contributing funders type since the main universities of these countries contribute to a large proportion of the national output and are acknowledged in their scientific publications.

These findings also show that most important funders in Middle East countries such as Saudi Arabia, UAE, Oman, Bahrain, and Jordan are universities by contrast to North African countries. Universities in the Middle East are granted direct government funds to support research initiatives in an independent way, however research funding is regulated through national calls for research projects in North Africa as confirmed previously by Currie-Alder (2015) and Hanafi and Arvanitis (2015).

**Table 5. Top 3 funders contributing to at least 1% of each country's total number of international publications in WoS in MENA (2008-2021)**

| CU | Funder | IPWFA | %IPWFA | FCU | Type |
|---|---|---|---|---|---|
| AFG | Japan Society for the Promotion of Science | 82 | 4.6 | JPN | Government |
| | UK Research & Innovation | 66 | 3.7 | GBR | Government |
| | United States Agency for International Development | 57 | 3.2 | USA | Government |
| ARE | National Science Foundation | 1,826 | 3.8 | USA | Government |
| | UK Research & Innovation | 1,785 | 3.8 | GBR | Government |
| | National Natural Science Foundation of China | 1,419 | 3 | CHN | Government |
| BHR | UK Research & Innovation | 167 | 3.7 | GBR | Government |
| | CGIAR, France | 64 | 1.4 | FRA | Research Institute |
| | National Institute of Health Research of Iran | 64 | 1.4 | IRN | Health |
| DJI | Agence Nationale de la Recherche | 20 | 8.3 | FRA | Government |
| | Center for Research and Studies of Djibouti | 11 | 4.6 | DJI | Research Institute |
| | UK Research & Innovation | 10 | 4.1 | GBR | Government |
| DZA | Ministry of Higher Education and Scientific Research, Algeria | 2,346 | 7 | DZA | Government |
| | Spanish Government | 977 | 2.8 | ESP | Government |
| | European Commission | 797 | 2.3 | EU | Government |
| EGY | King Saud University | 6,564 | 5.3 | SAU | Academic |
| | National Science Foundation | 4,108 | 3.3 | USA | Government |
| | National Natural Science Foundation of China | 4,094 | 3.3 | CHN | Government |
| IRN | Iran National Science Foundation | 3,748 | 2.5 | IRN | Government |
| | National Science Foundation | 3,642 | 2.4 | USA | Government |
| | UK Research & Innovation | 3,635 | 2.4 | GBR | Government |
| IRQ | UK Research & Innovation | 1,397 | 6.4 | GBR | Government |
| | Ministry of Higher Education and Scientific Research of Iraq | 1,179 | 5.4 | IRQ | Government |
| | Higher Committee for Education Development in Iraq | 489 | 2.2 | IRQ | Government |
| JOR | Jordan University of Science and Technology | 789 | 3.5 | JOR | Academic |
| | National Institutes of Health | 611 | 2.7 | USA | Government |
| | University of Jordan | 606 | 2.7 | JOR | Academic |
| KWT | Kuwait University | 683 | 5.7 | KWT | Academic |
| | National Institutes of Health | 488 | 4 | USA | Government |
| | Kuwait Foundation for the Advancement of Sciences | 475 | 3.9 | KWT | Nonprofit |
| LBN | National Institutes of Health | 1,053 | 4.7 | USA | Government |
| | United States Department of Health & Human Services | 957 | 4.3 | USA | Government |
| | American University of Beirut | 732 | 3.3 | LBN | Academic |
| | UK Research & Innovation | 230 | | GBR | Government |
| LBY | Ministry of Higher Education and Scientific Research, Libya | 81 | 1.8 | LBY | Government |
| | Government of Libya | 63 | 1.4 | LBY | Government |
| MAR | National Science Foundation | 2,188 | 7.9 | USA | Government |
| | Spanish Government | 2,066 | 7.5 | ESP | Government |
| | Centre National pour la Recherche Scientifique et Technique Morocco | 1,887 | 6.8 | MAR | Government |
| OMN | Sultan Qaboos University | 756 | 5.7 | OMN | Academic |
| | UK Research & Innovation | 483 | 3.6 | GBR | Government |
| | The Research Council Oman | 451 | 3.4 | OMN | Research Council |
| PAK | National Natural Science Foundation of China | 11,255 | 10.2 | CHN | Government |
| | Higher Education Commission of Pakistan | 10,300 | 9.4 | PAK | Government |
| | UK Research & Innovation | 4,495 | 4.1 | GBR | Government |
| PSE | Ministry of Science and Technology, China | 515 | 11.8 | CHN | Government |
| | UK Research & Innovation | 477 | 11 | GBR | Government |
| | National Science Foundation | 413 | 9.5 | USA | Government |
| QAT | Qatar National Research Fund | 5,324 | 17.2 | QAT | Nonprofit |
| | Qatar Foundation | 3,529 | 11.4 | QAT | Nonprofit |
| | National Natural Science Foundation of China | 2,206 | 7.1 | CHN | Government |
| SAU | King Saud University | 21,163 | 11.1 | SAU | Academic |
| | National Natural Science Foundation of China | 9,722 | 5.1 | CHN | Government |
| | King Abdulaziz University | 7,833 | 4.1 | SAU | Academic |
| SYR | United States Department of Health & Human Services | 91 | 2.6 | USA | Government |
| | National Institutes of Health | 89 | 2.5 | USA | Government |
| | UK Research & Innovation | 88 | 2.5 | GBR | Government |
| TUN | Ministry of Higher Education and Scientific Research, Tunisia | 4,332 | 9.6 | TUN | Government |
| | Spanish Government | 1,219 | 2.7 | ESP | Government |
| | European Commission | 1,130 | 2.5 | EU | Government |
| TUR | Turkiye Bilimsel ve Teknolojik Arastirma Kurumu | 15,692 | 11.7 | TUR | Government |
| | National Science Foundation | 9,223 | 6.9 | USA | Government |
| | UK Research & Innovation | 6,820 | 5.1 | GBR | Government |
| YEM | King Saud University | 293 | 5.3 | SAU | Academic |
| | UK Research & Innovation | 119 | 2.1 | GBR | Government |
| | National Natural Science Foundation of China | 111 | 2 | CHN | Government |

Also, the location of the funders allows to distinguish countries dominated by domestic or foreign funders. Most countries in MENA have at least one major domestic funder contributing to international publications. Afghanistan, UAE, Bahrain, Egypt, Egypt, Palestine, Syria and Yemen are the exceptions. In the case of Egypt and Yemen, King Saud University in Saudi Arabia appears as a main funder. This might be due to non-Saudi researchers who are affiliated to Saudi institutions and list them as affiliations (Bhattacharjee, 2011) but also as funders.

The United States of America, the United Kingdom, China, Japan, South Korea. Malaysia and some European countries such as France, Germany or Spain, are the locations of the main foreign funders. This suggests some relatively strong funding ties between specific countries and the MENA region. Saudi Arabia and Iran are the only countries in MENA appearing as a foreign location of one funder: King Saud University is the main funder in the case of Egypt and Yemen. Similarly, the National Institute of Health Research of Iran appears as a main funder located in MENA for Bahrain.

## 3.3. Location and type of major funders

To characterize further the funding landscape in MENA, the analysis of the location of all the major funders allows us to determine the geographical source of the funding granted to research projects in which researchers in the MENA region participated. Several distinctions can be made between domestic and foreign funders, but also between MENA and non-MENA funders. The contribution of all the funders by location is shown in Figure 5 for each country in MENA based on the share of publications where their name appears in the funding acknowledgments.

When considering the international publications, Figure 5 shows that several countries in MENA, such as Palestine, Afghanistan, Iraq, Libya, Algeria, Morocco and Bahrain have a significant share of contribution of foreign funders not located in MENA (top chart in Figure 5). This might be due to the presence of foreign funders acknowledged by international co-authors in the related publications. It is also possible that some foreign funders have specific funding programs focused on collaboration with researchers located in MENA.

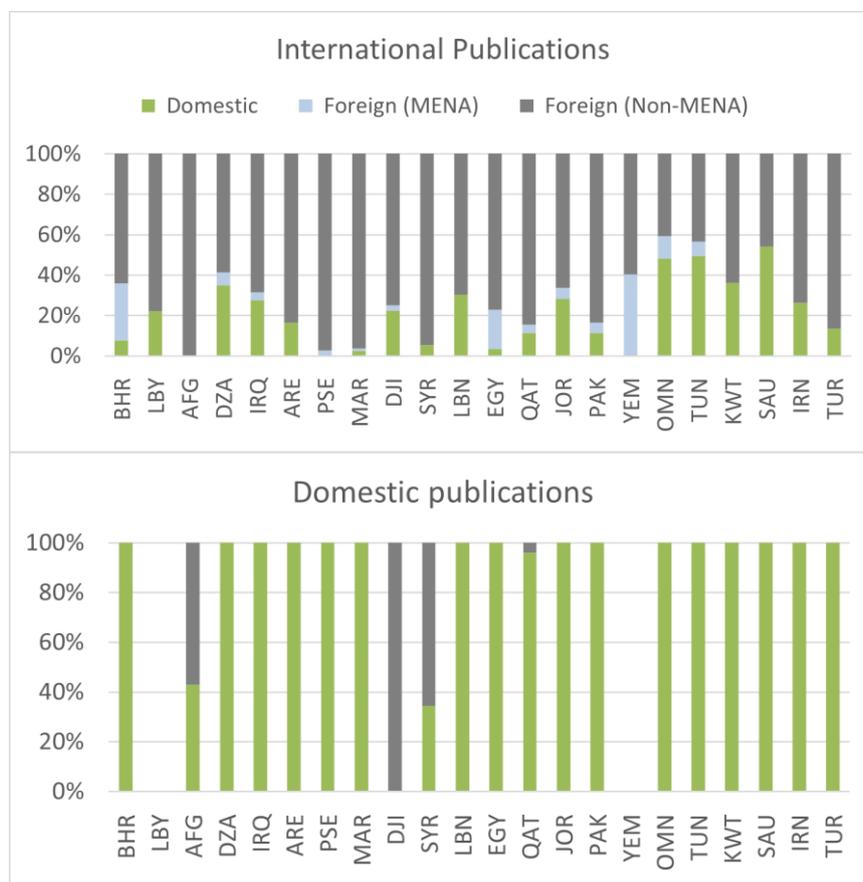

**Figure 5: Share of PWFAs of major funders in WoS with international co-authors (top) and only domestic authors (bottom) broken down by location type for each country in MENA (2008-2021).**

Countries like Tunisia, Oman, Jordan, Kuwait, Saudi Arabia, Pakistan, and Yemen show a higher involvement of both domestic and foreign funders (top chart). Turkey and Iran are dominated by one domestic funder as listed previously in Table 4.

When focusing only on domestic publications, most MENA countries show an exclusive contribution of domestic funders (bottom chart). However, researchers located in a few MENA countries such as Djibouti, Syria and Afghanistan acknowledged foreign funders in a substantial share of their publications. Also, it is worth noting that Libya and Yemen do not show any major contribution of funders in their domestic publications.

It seems the distribution of funders by location has become diverse in recent years for specific countries in MENA. However, the level of diversity in funding sources varies between MENA countries. These findings partially reveal that some differences exist between countries in MENA in terms of domestic funding capabilities and funding structures with some science systems relying more on foreign funding support than other economies especially when considering their international co-publications.

Another characteristic of funders considered in this section is their type. In Figure 6, the contribution of funders based on the share of publications where they are acknowledged is broken down by type for each country in MENA. As for the top 3 funders shown in Table 4, the government agencies seem to be the most active in the region in terms of research funding. Then, the academic funders, the nonprofit funding organizations and the research councils follow. Figure 6 also shows that some countries with a relatively high number of funders such as Qatar, Palestine, and Morocco have a contribution to international publications from a diverse range of agency types (right of the top chart). The chart also highlights the strong presence of government funders in all MENA countries and academic funders in most countries.

However, when focusing only on the domestic publications (bottom chart), the types of the funders are less diverse than in the international publications. The government and academic sectors appear to be the major contributors in terms of publications with funding acknowledgments. In most countries in MENA, the government and/or academic sector are even the sole major source of funding acknowledged in domestic publications. However, a few countries such as Oman, Kuwait, Qatar but also Egypt and Lebanon still show other types of funders including research institutes, nonprofit and research councils being also main domestic sources of funding.

These results are aligned with statistics reported by UNESCO (2015). The bulk of Global Expenditure on Research and Development (GERD) in many countries in MENA is reported to be performed mainly by the government sector, followed by the higher education sector. At the time of the report, the UNESCO also highlights that the private sector plays a small or no role in the research activities. However, some exceptions such as Jordan, Morocco, Oman, Qatar, Tunisia and the UAE where the private sector did contribute on average 25% of GERD. But such contribution is not necessarily reflected in the funding acknowledgments found in scientific publications as per the results of this study. This may suggest that these funding contributions are directly not visible in the scientific publications as the final outputs but may support other purposes such as funding of patenting activities or commercialization efforts. Also, the funding support provided to research and development activities may also simply be reflected in the affiliations of the authors of a scientific publication.

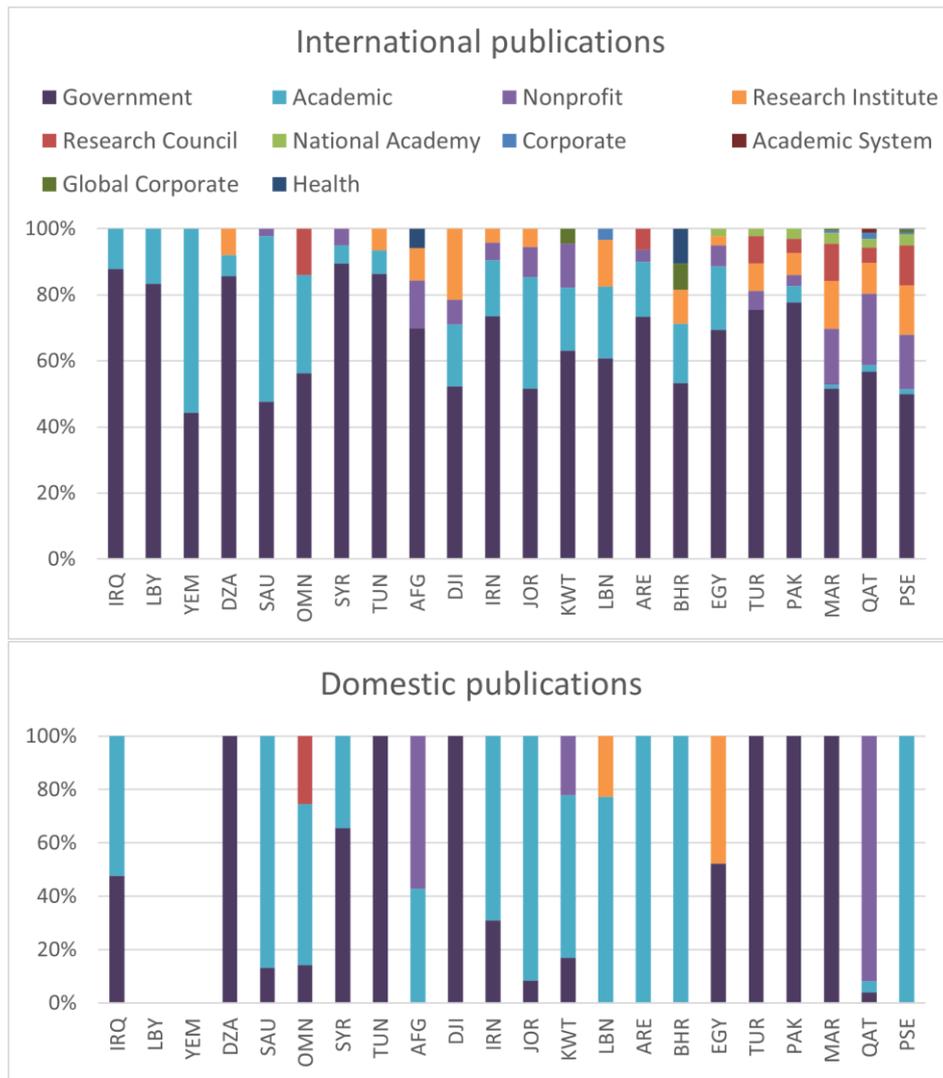

**Figure 6: Share of PWFAs of major funders in WoS with international co-authors (top) and only domestic authors (bottom) broken by agency type for each country in MENA (2008-2021).**

### 3.4. Qualitative analyses of funding acknowledgments in international publications

To complement the previous results, several qualitative analyses of funding acknowledgments found in distinct random samples of international publications were conducted.

Based on the previous scientometric analyses, the clarity of research funding remains a challenge when acknowledged in international publications. It is not very clear who the funding recipient is. The first random sample (S1) is composed of 100 random international papers with funding acknowledgments. The purpose of analyzing this sample is to gain deeper insights into the context of the funding reported in such international publications.

Table 4 and 5 highlight the significant presence of universities in funding acknowledgments. In many instances, universities seem to play a funder role. Notably, in domestic publications (Table 4), universities appear to be prominent funding contributors. The second random sample (S2) contains 25 random papers with at least one university acknowledged as a funder. The aim of the analysis of this sample is to clarify the funding structure when a university is acknowledged in the funding statement of a scientific publication.

These two random samples were extracted from the in-house version of the Web of Science T-SQL database held at the Centre for Science and Technology Studies-CWTS (Leiden University). Random sampling SQL queries with relevant clauses were run to obtain the random samples (S1: 100 IPWFAs with at least an author from MENA. S2: 25 IPWFAs with at least one university acknowledged as a funder). The MENA countries these two random samples correspond to are shown in Figure 7.

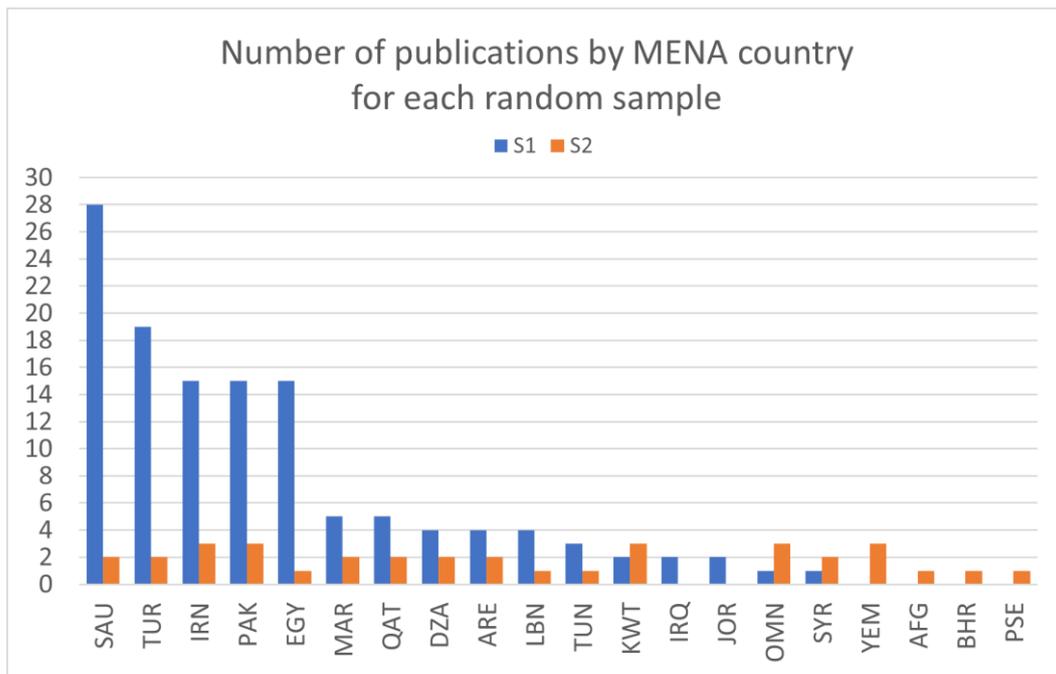

**Figure 7: Number of publications by MENA country for each random sample.**

The funding acknowledgments found in the papers of these two random samples were analyzed manually. The data was structured based on groups of recurring themes relevant to the research questions of this study.

3.4.a. A qualitative analysis of funding acknowledgments in 100 random international publications

Several elements were analyzed in the first random sample. Findings are reported in Table 6. The first element of this analysis consists of analyzing whether the funding is attributed to specific author(s). 30% of the international publications with funding acknowledgments of this random sample contain information about the attribution of the reported funding support. In most cases, authors simply juxtapose the funders from which they receive funding and do not mention who received the funding. In about two thirds of publications with attributable funding, the funding is granted to the international co-author.

Then, the second aspect of the manual analysis concerns the acknowledgment of the authors affiliations. About half of the publications show an acknowledgment of at least one of the affiliations of the authors, i.e., the employers of the researchers, and 46% of the publications contain the grant details of the funding support received by the authors.

Next, the locations of the acknowledged funders were also analyzed from the perspective of the MENA country for which the publication is counted. It is worth reminding that the IPWFAs are fully counted against the location of each acknowledged funder that can be of mixed origin, with a distinction between domestic funding (the funder shares the same location as the MENA country for which the publication is counted) and foreign funding (the funder is located in a foreign country in MENA or Non-MENA). The full counting of IPWFAs by funder location explains why the total percentage by funder location exceeds 100%. In this first random sample, the share of foreign funders from MENA and non-MENA countries is 79% and 9% respectively. However, the proportion of domestic funders is 42%.

**Table 6. Proportion of PWFAs by funder location, funder type, attribution of funding and acknowledgment of author affiliations in the first random sample**

|  |  | *First random sample* |
|---|---|---|
|  | *Proportion of papers* | *%IPWFA* |
|  | Funding attributable to specific author? | 30% |
|  | Affiliation(s) Acknowledged? | 54% |
|  | Grant(s) details | 46% |
| *Funder Location* | Foreign Non-MENA | 79% |
|  | Foreign MENA | 9% |
|  | Domestic | 42% |
| *Funder Type* | Government | 64% |
|  | Academic | 47% |
|  | Nonprofit | 20% |
|  | Corporate | 11% |
|  | Research Institute | 9% |
|  | Research Council | 5% |
|  | National Academy | 5% |
|  | Global Corporate | 3% |
|  | Health | 1% |
|  | Academic System | - |
|  | Partnership | - |

Similarly, funders acknowledged in publications can be of mixed type and IPWFAs are fully counted towards each type. This also explains why the total percentage of IPWFAs by funder type exceeds 100%. This first random sample shows a relatively high variety of funders in terms of types. Government funders represent the highest share (64%), followed by academic (47%) and nonprofit ones (20%).

Overall, these findings confirm the results of Section 3.2 found with regards to the locations and types of the acknowledged funders. Additionally, a substantial share of publications contains funding acknowledgments to the authors' employers. However, the funding is attributed to specific authors in a lesser extent. In these international publications with attributable funding, about two thirds of them show that the funding is granted to the international co-author.

3.4.b. A qualitative analysis of 25 random publications with at least one university acknowledged as a funder

Like Table 6, Table 7 lists the different elements analyzed in the second random sample consisting of 25 random publications with at least one university acknowledged as a funder. In this sample, the funding is explicitly attributed to a specific author in 24% of the publications. In most cases, authors simply juxtapose the funders from which they receive funding without explicit attribution. Nevertheless, all the publications in this sample contain an acknowledgment of the university with which at least one of the authors is affiliated.

This sample consists of publications which mention at least one university as a funder which explains the 100% value for the proportion of publications with a funder of academic type. About one third of the publications of this sample also acknowledge a government funder. Global corporates are also acknowledged but in a much lesser extent (4%).

**Table 7. Proportion of PWFAs by funder location, funder type, attribution of funding and acknowledgment of author affiliations in the second random sample**

|  |  | *Second random Sample* |
|---|---|---|
|  | *Proportion of papers* | *%IPWFA* |
|  | Funding attributable to specific author? | 24% |
|  | Affiliation(s) Acknowledged? | 100% |
|  | Grant(s) details | 48% |
| *Funder Location* | Foreign Non-MENA | 76% |
|  | Foreign MENA | 8% |
|  | Domestic | 72% |
| *Funder Type* | Government | 32% |
|  | Academic | 100% |
|  | Nonprofit | - |
|  | Corporate | - |
|  | Research Institute | - |
|  | Research Council | - |
|  | National Academy | - |
|  | Global Corporate | 4% |
|  | Health | - |
|  | Academic System | - |
|  | Partnership | - |

In terms of locations of the acknowledged funders, the share of foreign funders (MENA and non-MENA) in the first and second random samples are similar. However, the proportion of domestic funders is substantially higher in the second sample than in the first one, respectively 72% and 42%. In fact, in the second random sample, all the MENA authors acknowledge their employers, i.e., the universities where they work, as their funding providers. In most cases, the acknowledgments found in that sample consist of giving credit to the authors' employers without specific details. In such instances, the authors may simply refer to the salaries they receive from their employers to conduct research. However, other funding acknowledgments provide more details in terms of received funding support (awards of grants with grant details, purchase of equipment, travel exchange programs, special projects, etc.).

### 3.5. Survey of 220 corresponding authors of IPWFAs

In order to contextualize the previous results, this section presents the findings of the survey of 220 corresponding authors of 220 IPWFAs constituting the third random sample (10 publications for each country in MENA). The corresponding authors of these papers were contacted by email with a series of open and closed questions related to the funding reported in their publications. Each email was personalized, featuring title and a direct link to the related IPWFA but also the list of acknowledged funders. The primary objectives of these questions were to determine whether:
- the reported funding was dedicated to a specific research project, or if the funding was internal (e.g., salaried research time),
- the funding was shared among co-authors,
- and whether there were any requirements regarding international collaboration with specific countries.

28 responses were received, representing a response rate of 12.7%. Response rates of surveys vary based on multiple factors such as the employed survey method, the demographics of the surveyed population, or the field (Langfeldt et al., 2023). However, 12.7% is higher than what is observed in a similar online email survey of researchers (Ni & Waltman, 2023).

The countries of affiliation of the 28 corresponding authors of the IPWFAs, at the time of publication, are listed in Table 8. MENA countries are marked with an asterisk symbol. It is worth reminding that the corresponding authors of international publications were not necessarily from MENA at the time of publication of the IPWFAs under study and that some corresponding authors had multiple affiliations from different countries.

**Table 8. Distribution of the 28 respondents by country of affiliation**

| Country | Number of respondents | Country | Number of respondents | Country | Number of respondents |
|---|---|---|---|---|---|
| USA | 4 | Belgium | 1 | Libya (*) | 1 |
| Pakistan (*) | 3 | Canada | 1 | Malaysia | 1 |
| Turkey (*) | 2 | Djibouti (*) | 1 | Palestine (*) | 1 |
| Switzerland | 2 | England | 1 | Pakistan (*) | 1 |
| Oman (*) | 2 | Ethiopia | 1 | Saudi Arabia (*) | 1 |
| France | 2 | India | 1 | Syria (*) | 1 |
| UK | 1 | Japan | 1 | Tunisia (*) | 1 |
| Austria | 1 | Lebanon (*) | 1 | | |

Quotes are reported in this section to illustrate the findings. These responses were grouped under three topics:
- the distinction between funding received for a dedicated research project and the internal funding or salary received from the author's employer to do research,

- the distinction between research projects where the funding is shared among all co-authors and the juxtaposition of research funding acknowledged in publications,
- and requirements in terms of international collaboration.

*Dedicated funding versus salaried research time*

Out of the 28 responses, 10 authors clarified that the funding support they received was dedicated to the research reported in their paper. Several answers provide insights into the distinction between research funding dedicated to specific projects and the internal funding or salary received from the authors' employers to conduct research. The authors' statements emphasize the source and purpose of the funding they received, clarifying the nature of financial support for their research activities.

In the first two quotes below, the funding is explicitly described as being provided for dedicated research projects. The mention of the French Ministry of Foreign Affairs and the University Grants Commission highlights the origin of these funds, which were granted for specific research initiatives. This indicates that the research was intentionally supported by external and specific entities, underlining the clear distinction between these funds and internal resources. Here, the funding is attributed through PhD programs.

> The first author was a PhD student at that time and these funds were granted from the French ministry of foreign affairs.

In the second quote below, the author also mentions specific programs with the pharmaceutical industry. This may suggest a close collaboration between the author's university and the pharmaceutical company which funds specific research projects.

> Yes, it was a funded project. I was a Ph.D. student on this project. This project was granted by UGC [University Grants Commission] to my supervisor. This research was funded through pharmaceutical research programs.

The following quote illustrates how dedicated funding was provided for investigating a specific topic, in this case, waterpipe tobacco smoking. The grant was awarded by a major funder, the US National Institutes of Health, aligning the financial support directly with the research subject. This quote also emphasizes the targeted nature of funding for research purposes.

> The two US National Institutes of Health grants (one from the National Institute on Drug Abuse and the other from the National Cancer Institute) that were cited in the publication were both awarded for the investigative team to study waterpipe tobacco smoking, which is the subject of the publication. So, yes, this funding was provided (in part) and "dedicated" for the purpose of studying the questions that were addressed in this publication. We were also funded to address other questions on the same topic (waterpipe tobacco smoking)

The final quote provides an example where the research did not receive dedicated funding. Instead, the authors' institutions were contracted to provide technical support, leading to collaborative publications resulting from this collaboration. This scenario demonstrates how research outputs can arise from support work, even when the funding is not explicitly dedicated to the research at hand. The author also clearly mentions that, in some cases, scientific publications are expected to be produced from certain types of technical and research works.

> We did not receive dedicated funding for the research in this paper. Johns Hopkins Bloomberg School of Public Health (JHSPH) and the Indian Institute of Health Management Research (IIHMR) were contracted by the Ministry of Public Health of Afghanistan as a third party to support monitoring and evaluation of

> their basic and hospital package of services. However, this work resulted in many joint publications between JHSPH and IIHMR researchers and MoPH officials. Institutions such as Hopkins pretty much expect publications to result from all technical support work.

Overall, these quotes clarify the significance of understanding the source, purpose, and intentions behind research funding. The distinction between funding dedicated to specific projects and internal resources is crucial for accurately assessing the financial landscape of scientific research.

*Shared funding among co-authors or juxtaposition of research funding?*

Out of 28 responses, 4 authors mentioned that the received funding was shared among the co-authors. Several answers from the corresponding authors allows to make the distinction between research projects where funding is shared among all co-authors and projects where each researcher reports their individual funding. Their answers emphasize how funding is allocated, acknowledged, and used within collaborative research efforts.
The first two quotes are instances where the funding is distributed among co-authors. In the first quote, the authors highlight that some co-authors received salaries from the reported funding, underscoring the shared nature of the financial support.

> A couple of co-authors received salary from the reported funding to carry out the studies.

The second quote indicates that salaries for multiple authors were drawn from the combined funding sources, reinforcing the collaborative aspect of the project.

> This paper was the result of a collaboration that was supported by a grant that I received from the Swiss National Science Foundation, as well as funding from the Leenards and Jeantet Foundations. Salaries for 7 of the authors were drawn from these fundings.

In contrast, the acknowledged funding is not shared by co-authors in most cases. The authors tend to mention and use independently the funding they receive. the subsequent quotes reveal situations where acknowledged funding is not shared among co-authors. Authors simply mention and utilize individual funding independently, indicating a non-shared funding approach.

> Maziak, Rastam, Ibrahim and Ward were funded by (and salary supported by) DA024876, while Shihadeh was funded by (and salary supported by) CA120142). Eissenberg was funded by (and salary supported by) both DA024786 and CA120142.

This approach seems more frequent in cases where co-authors have distinct funding sources, as showcased in quotes where different co-authors used different funding sources based on their country of origin.

> Different co-authors used different funding sources. Netherlands' and Turkish funding was granted to the authors from The Netherlands and Turkey.

The two quotes below provide context into projects where external funding is not the primary factor behind collaboration. In one case, the research was not externally funded, and the authors leveraged their own affiliations for acknowledgment.

> This was not externally funded. The study was run by a PhD student (who was externally funded by the Libyan government to do her PhD) and myself (fully

> funded by the University of Liverpool). We could have described it as 'Funded by the Libyan Government' – but they funded the student to do the PhD, not the clinical trial. So, we just used University of Liverpool as that is my university and the place of the PhD.

In another, the collaboration is driven by shared research interests rather than shared funding.

> I knew Victor and Wissem much earlier than the time when we wrote the paper. The fact we share mathematical research interest is the only reason why we collaborated. There is no special shared funding relating to this article/work. The mentioned grants were used by the specified authors only.

These quotes collectively highlight the diverse approaches to acknowledging and sharing research funding among co-authors. While some collaborations involve shared financial resources, others rely on individual funding sources. This variability reflects the complexity of funding dynamics within collaborative research projects, where both shared and non-shared funding models play a role in driving scientific research activities.

*International collaboration requirements*

The survey of the corresponding authors also offers insights into the requirements imposed by funders for international collaborations in specific research projects. Their answers illustrate how the support provided by funders can influence the nature and extent of collaboration between researchers from different countries.

The first quote below points to a collaboration facilitated by the supervisor of a PhD student. This suggests that the collaboration was likely influenced by the co-author's institutional connection, showcasing how affiliations with specific organizations can drive international collaboration.

> The co-author was my PhD advisor who worked in Intergovernmental Authority on Development (IGAD), Djibouti

In the following quote, the mention of a postdoctoral funded by the Higher Education Commission in Pakistan and completed in Japan highlights the role of funding from a specific country in fostering international scientific mobility and collaborations for advanced research positions.

> This was a Pakistan HEC post doc funding which was completed in Japan.

The third quote below reveals a certain commitment and intentions to geographical diversity and capacity building. For instance, the following quote is related to a project funded by internal budgets that emphasizes the importance of Global North-South scientific collaboration.

> It [the funding] was only used for the PhD grant and the PhD student travels between the two involved countries, France and Tunisia. the reason for [my] collaboration with a colleague from Tunisia was to promote North-South scientific collaboration, training young south countries students. As a French researcher, I am particularly involved in such cooperations.

The subsequent quotes point to instances where internal budgeting and affiliations drive collaborations. In these cases, the funding sources may not be tied to external expectations for international collaboration but enable opportunities for researchers to engage across borders.

> This research projected was supported by the Center's internal budget. It was internal research funding. The colleague at the time of the research was affiliated to our institution and had assumed a position at Qatar University.

Another example specifies that temporary or visiting positions also play a critical role in funding support and funding acknowledgment:

> I visited the biology department at Sultan Qaboos university [which funded the research project], there was an opportunity to collaborate on a small project.

In a different scenario described below, collaboration arises informally due to shared interests and complementary scientific disciplines. The collaboration is driven by the mutual benefits each researcher brings to the project, highlighting how rigor in research is promoted through transdisciplinary teamwork.

> We are all close friends working from complementary scientific disciplines (Engineering, Epidemiology, Medicine, Psychology). We continue to work together in a collegial and transdisciplinary fashion today. We work together because each of us makes the work more rigorous, more revealing scientifically, and more fun.

In addition to specific required expertise, the facilities or equipment available in some research institutions also play a role in shaping collaborations and support acknowledged by authors. For instance, in the following quote, the author mentions that the co-author was instrumental in doing specific sample analyses that the corresponding author would not have been able to do in his own institution.

> The funding was provided by the university [Izmir Institute of Technology], and it was used solely to finance the required chemicals and sample analysis. There were no conditions in terms of international collaboration. The colleague from Germany was helpful in some key sample analysis.

Finally, a corresponding author explains that the international collaboration with researchers from specific countries was a requirement from a funder supporting the research project.

> Yes, this funding was dedicated to this research project. EURAPMON and Greater Los Angeles Zoo Association are the funders. And this funding was also used by co-authors. It was required from ERAPMON to collaborate with colleagues from specific countries [Djibouti, Oman] but there was no such a requirement from GLAZA.

Overall, these quotes showcase a range of international collaborations driven by various funding sources and motivations. Whether it is funders' requirements, institutional affiliations, specific funding programs, or personal connections, funders also play a crucial role in shaping the global collaborative landscape in scientific research.

## 4. Discussion and conclusions

The aim of this study was to better understand the funding structure of scientific research in the Middle East and North Africa (MENA) region based on the funding acknowledgments found in scientific publications. An important element to consider when analyzing funding acknowledgments is the distinction between domestic publications and international publications. The recent increase of international co-authorship in scientific publications likely explains why some countries in MENA show a relatively high level of contribution of foreign funders. These foreign funders may have financially supported researchers who have co-authored scientific publications with researchers located in MENA. Also, the analysis of domestic publications reveals the contribution of the main domestic funders but also the role some foreign funders play in funding domestic research.

First, this study shows that the proportions of publications with funding acknowledgments vary greatly across MENA countries. Overall, countries in MENA have lower proportions of publications with funding acknowledgments than the World. Saudi Arabia and Qatar top the list with about half of their publications containing a funding acknowledgment. This may demonstrate the high level of funding available at a country level, but one needs to remember that some factors might impact what authors acknowledge in their publications. There are many reasons that influence the inclusion or exclusion of acknowledgments in scientific publications. For example, the presence of funding acknowledgments might be a requirement of the funder which supported financially the authors (Alvarez & Caregnato, 2018). In collaborations for multicentric medical research, minor contributors who do not qualify for authorship are often recognized with acknowledgments because of some journals' editorial policies limiting the number of authors per article (Alvarez & Caregnato, 2021). Researchers paid by their employer such as a public university to do research, may be considered to be funded by the government but this would not necessarily be reported as a funding acknowledgment in their publications. In this case, the researcher's affiliation may better reflect the support of their research institution as a funder.

The share of publications with funding acknowledgments follows an uptrend across all MENA countries except in countries which have witnessed conflicts or unrests in the recent years. Next, the names of the funding organizations found in the funding acknowledgments of scientific publications indexed in Web of Science (WoS) were unified. This unification process reveals a diverse funding landscape with Iran, Turkey, Saudi Arabia and Egypt having the largest number of domestic funding organizations in the region. The difference in terms of number of funders might be explained by the size of the respective country but also by the different national funding structures.

Three groups of countries in MENA can be distinguished as per their number of funders: when considering all their publications, some countries like Qatar, Palestine and Morocco are dominated by several dozen funders contributing to at least 1% of the total number of national publications; then other countries such as Saudi Arabia, the UAE, Kuwait or Pakistan rely mainly on a few funders. Finally, countries like Turkey or Iran are dominated by one or two funders. However, when analyzing only the domestic publications, then the findings reveal only a few major domestic funders for most countries in MENA.

The findings also show the contribution of domestic and foreign funders to the scientific output of each country. Some countries, like Turkey, Iran, Oman, Kuwait, Jordan and Saudi Arabia, have a relatively high level of contribution of domestic funders mentioned in their scientific papers. In contrast, other countries in MENA seem to rely more on foreign funding sources.

The main funders found in scientific publications in MENA are government and academic organizations. This is not surprising since most of the scientific research in MENA is conducted by public universities funded by the governments and more specifically by the Ministry of Higher Education and Research or equivalent national institutions. Nevertheless, this study also shows the funding contributions of other sectors or types such as nonprofit institutions, research councils or national academies. Corporations appear as funders mainly in international scientific publications, but their contribution is much less visible. Public-private partnerships, and more specifically collaboration between the public and the private sector, have gained traction in scientific research funding in recent years and help to accelerate the translation of research into practical applications.

The different qualitative analyses allow a better understanding of the funding reported by authors in funding publications. More specifically, they clarified when the funding was attributed to specific authors, whether the funding was internal or external, and if it was dedicated to a specific project. Furthermore, the responses received from the surveyed corresponding authors shed light on collaborations observed in international co-authored publications. These analyses contribute to the limited body of qualitative studies on acknowledgments in science from the perspective of researchers (Alvarez & Caregnato, 2020; Cronin & Overfelt, 1994; Roa-Atkinson & Velho, 2005). They confirm the important role played by acknowledgments in the primary communication process and also contribute to the literature about the operationalization of funding (Aagaard et al., 2021). The attribution and the amalgamation (fusion or juxtaposition) of funding are revealed by the employed qualitative methods and provide context on the collaboration between co-authors (Alvarez & Caregnato, 2020).

This study also contributes to a growing literature on research funding (Gök et al., 2016) which concerns various stakeholders: governments, public and private funding institutions, universities, research managers, and researchers. Several MENA countries have made investments in science and technology capacity to promote research and innovation (Schmoch et al., 2016; Shin et al., 2012; Siddiqi et al., 2016). Policy makers also changed their research governance (Currie-Alder, 2019). For instance, Arab countries have set national research priorities and changed the way to access public funding. Research funds expanded over the past two decades (Currie-Alder, 2015) and incentives were set to collaborate with distant and scientifically proficient foreign partners to connect with global scientific networks and improve the rank or Arab research organizations in international rankings of academic quality (Currie-Alder, 2019; Currie-Alder et al., 2018).

This present study sheds light on the recent trends on the landscape of funding in scientific research conducted in MENA. While the findings are significant, the analysis of funding acknowledgments comes with limitations which may underestimate the role of institutional funding which researchers do not always mention explicitly in their acknowledgments. In fact, researchers also mention the institutional funding their received by reporting their institutional affiliations in their scientific publications. Although some limitations exist, this study provides insights into the structure of scientific funding in MENA especially in terms of funding source, type and funding trends. These insights can be used by policy makers to monitor and design their research funding programs (Alvarez-Bornstein & Montesi, 2021; Paul-Hus et al., 2016; Wang & Shapira, 2015). For instance, from a funding policy and research evaluation perspective, policymakers could examine whether researchers in specific countries who received funding have produced any scientific publications (Albrecht et al., 2009; Álvarez-Bornstein et al., 2018; Alvarez & Caregnato, 2018, 2021; Costas & Yegros-Yegros, 2013; Díaz-Faes & Bordons, 2014; Gao et al., 2019; Gök et al., 2016). Another example may consist of determining the trends in financing a specific subject or research area has received (Dorsey et al., 2006). Other works using funding acknowledgments have involved the mapping of funders to specific fields (Lewison & Dawson, 1998; Lewison et al., 2001) or specific funding programs (Boyack & Börner, 2003). Rangnekar (2005) also conducted an analysis of the mention of the Multiple Sclerosis Society, as a funder, within multiple sclerosis-related publications to analyze its visibility, its research orientations as well as its impact.

Finally, the results of this study open doors to new research opportunities. This study also reveals that funding acknowledgments data is a valuable starting point to understand funding structures and funding mechanisms. For instance, future work could focus on transnational research funding and how different forms of funding are used and reported by researchers. More specifically, future research may seek to analyze the collaboration and co-funding networks by country at the research area level, which may provide useful insights into the structures of funding mechanisms with regards to collaboration and research fields. The recent expansion of grant data coverage through the integration of Pivot-RP data in WoS[2] means that more records in WoS will have funding information for the first time and may also open doors to more granular analyses including the funding types (e.g. research, training, awards, collaboration agreement, travel, postdoctoral award, etc.). However, this study also reveals the complexity hidden in funding acknowledgments and the importance of both quantitative and qualitative methods to reveal information not explicitly communicated in funding acknowledgments. Ideally, to use funding acknowledgments quantitatively, authors will need to be more explicit about the funding context of their research. Scientific journals, conferences, books editors and preprint servers may have an important role to play in streamlining the funding information communicated by authors. Achieving a complete standardization of funding acknowledgments may not be feasible, necessitating an enduring need for qualitative analyses. Such quantitative and qualitative analyses are expected to better inform the policy makers on the funding structure of national science systems in MENA. They are also expected to provide insights on how the national science systems can best be funded in the near future.

**Data availability**
The research presented in this paper uses Web of Science data made available by Clarivate. The author is not allowed to share this data. However, the list of unified funders with their type and country is available in a data repository (El-Ouahi, 2023).

**Acknowledgments**
I would like to thank Ludo Waltman and Thomas Franssen for their insightful comments and valuable suggestions that greatly improved the quality of this paper. I extend my gratitude to two anonymous reviewers whose comments enhanced the quality of this article.

**Competing interests**
The author is an employee of Clarivate Analytics, the provider of Web of Science and InCites.

**Funding information**
This research project received no funding.

---

[2] https://clarivate.com/webofsciencegroup/release-notes/wos/web-of-science-release-notes-july-20-2023/